\documentclass[onecolumn]{IEEEtran}

\ifCLASSINFOpdf
  \usepackage[pdftex]{graphicx}
  \graphicspath{{images/},{media/}}
  \DeclareGraphicsExtensions{.pdf,.jpeg,.png}
\else
\fi

\usepackage{url}
\usepackage{textcomp}
\usepackage{hyperref}
\usepackage{booktabs}
\usepackage{tabularx}
\usepackage[utf8]{inputenc}
\usepackage[frozencache=true,cachedir=.]{minted}
\usepackage{float}
\usepackage{footnote}
\usepackage{subfigure}
\usepackage{color}

\begin{document}

{\color{red}\begin{flushright}Accepted for publication at IEEE Consumer Electronics Magazine on 22.04.2018 - published version may differ\end{flushright}}

\textbf{\large Designing a blockchain-based IoT infrastructure with Ethereum, Swarm and LoRa}

By Kaz{\i}m R{\i}fat \"{O}zy{\i}lmaz and Arda Yurdakul\newline

Today, the number of IoT devices in all aspects of life is exponentially increasing. The cities we are living in are getting smarter and informing us about our surroundings in a contextual manner. However, there lay significant challenges of deploying, managing and collecting data from these devices, in addition to the problem of storing and mining that data for higher-quality IoT services. Blockchain technology, even in today\textquotesingle s nascent form, contains the pillars to create a common, distributed, trustless and autonomous infrastructure system. This paper describes a standardized IoT infrastructure; where data is stored on a DDOS-resistant, fault-tolerant, distributed storage service and data access is managed by a decentralized, trustless blockchain. The illustrated system used LoRa as the emerging network technology, Swarm as the distributed data storage and Ethereum as the blockchain platform. Such a data backend will ensure high availability with minimal security risks while replacing traditional backend systems with a single ”smart contract”.

\IEEEpeerreviewmaketitle

\section{Introduction}
The Internet of Things is the backbone for creating smart buildings, smart energy systems, smart transportation and smart health care, which are the vital components of smart cities~\cite{mohanty2016everything}. In order to ensure safe and rapid adoption of IoT solutions, three essential aspects should be recognized: security, trust and identity of things~\cite{montgomery2015future}. Blockchain technology not only addresses these three concerns, but also shows a clear path for integrating all kinds of IoT devices to a common blockchain-based infrastructure as well~\cite{lee2017bidaas}~\cite{zorunluluk}. This approach defines a different role for every IoT device based on its capabilities and power requirements, therefore conforming with mobile-edge computing vision for consumer electronic devices~\cite{corcoran2016mobile}.

IoT deployments suffer from the problem of collecting, storing, and processing data in the cloud. An IoT platform should support multiple devices and services from different stakeholders, scale in a reliable and decentralized manner and offer tools and support for the rapid creation of applications and their execution~\cite{gubbi2013internet}. Selecting a unified method that enables data transmission from all kinds of IoT devices is another problem. In order to propose a solution, it is imperative to analyze what the future of the IoT landscape will look like. Ericsson predicts that low-power wide-area (LPWA) technologies like LoRa and Sigfox that operate in unlicensed band, and cellular-based NarrowBand IoT (NB-IoT) will be the great enablers for mass deployment of low-power end devices~\cite{ericsson2016cellular}. The current paradigm of short-range (NFC, Bluetooth, Zigbee), mesh-topology (wireless sensor networks) communication, which limits the coverage area of IoT devices, is challenged by the low-rate, long-range communication paradigm with star topology~\cite{vangelista2015long}. This shift in wireless communication technology may enable the deployment of low-power, low-cost devices with extended coverage in massive amounts. This gateway-centric approach inherently brings the possibility of implementing software solutions on IoT gateways.

\begin{minipage}{0.50\textwidth}
In 2008, Satoshi Nakamoto published the Bitcoin paper, which proposed a novel digital currency based on a decentralized, trustless infrastructure~\cite{nakamoto2008bitcoin}. All transactions are stored on a distributed database called {\it blockchain} and continuously verified using public-key cryptography by all peers in the system, thus eliminating the need for a central authority. Modifying contents of the chain without being caught is only possible with owning at least 33\% of the total computational power~\cite{DBLP:journals/corr/EyalS13}.
\end{minipage}
\begin{minipage}{0.55\textwidth}
  \begin{figure}[H]
    \centering
    \includegraphics[scale=0.45]{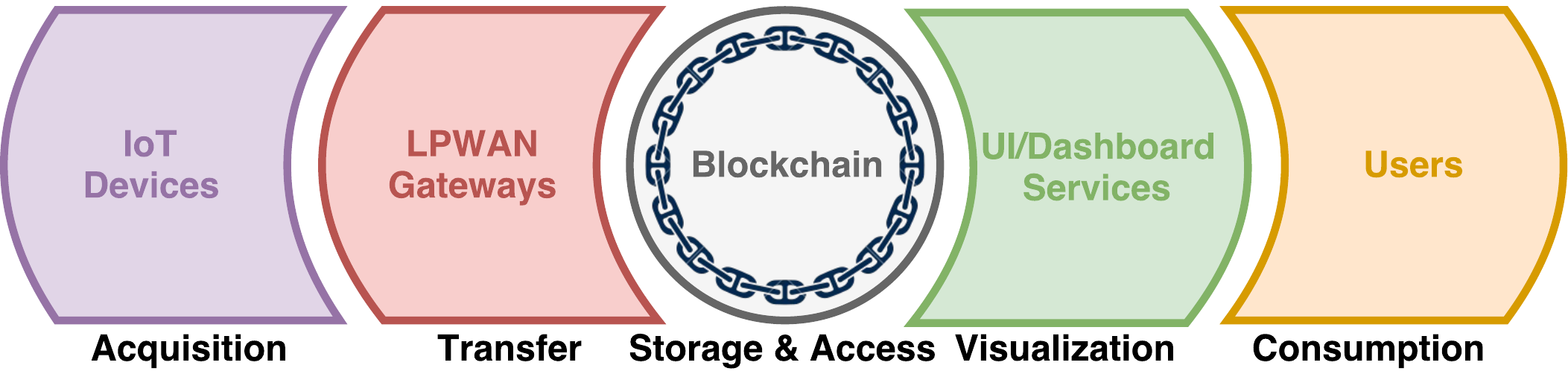}
    \caption{IoT System Overview}
    \label{fig:overview}
  \end{figure}
\end{minipage}

In this paper, a blockchain-based IoT infrastructure is described for the emerging, gateway-centric communication technologies that the majority of consumer electronic devices will use (Figure~\ref{fig:overview}). We also propose different methods of integration for various types of end devices. Our software solution aims to a) standardize the way IoT devices discover, communicate and send data to their data repositories, b) create a peer-to-peer, fault-tolerant and DDOS-resistant infrastructure for IoT deployments, c) facilitate a standard way to query and acquire IoT device data for the creation of next generation products and services.

In order to achieve these goals, we have investigated how a peer-to-peer network may be used to store data and code fragments, which in turn enables IoT gateways to push data and interact with other peers by means of a unified interface. As a proof-of-concept, a blockchain client is integrated to a LoRa gateway. A private (although not mandatory), peer-to-peer network, which makes use of these new blockchain-enabled LoRa gateways, is set up for demonstrative purposes. The peers in this network send data through the IoT gateway, store it in a torrent-like distributed file system, save handles of data chunks to blockchain, interact with events, and access uploaded data using a blockchain infrastructure.

\section{Low-Power Wide-Area Networks (LPWAN)}
\label{sec_LPWAN}
Low-power wide-area networks (LPWAN) is the new wireless connectivity that introduces star networks, as opposed to traditional wireless sensor networks properties like short-range, mesh networks. Most LPWAN technologies make use of low-power, low-cost end devices while covering distances over kilometers because of protocol and transceiver architecture efficiencies. It is possible to build inexpensive sensor nodes without SIM cards while having more robust gateways to connect and transfer data to IP-based networks.
Industrial representation of LPWANs is divided into two main categories: licensed and unlicensed band operation. Three technologies operating in the licensed band are eMTC (LTE Cat.M), EC-GSM and NB-IoT~\cite{3gpp2016progress}. Licensed technologies operate well in dense urban areas with good QoS. On the other hand, unlicensed technologies like LoRa and Sigfox provide generally better coverage, lower power and lower cost. Their downsides include a lower QoS and no guaranteed latency \cite{bardyn2016iot}. Table~\ref{tab:lpwa} shows the operating frequencies, bandwidth and data rate of these LPWAN technologies~\cite{nokia2016lteevo}.

\begin{table*}[htbp]
\centering
\caption{LPWA IoT Connectivity Overview}
\label{tab:lpwa}
\begin{tabular}{llllll}
\toprule
 & LoRa & Sigfox & NB-IoT (Rel.13) & eMTC (Rel.13) & EC-GSM (Rel.13) \\
\midrule
Range & \textless11km & \textless13km & \textless15km & \textless11km & \textless15km\\
Max coupling loss & 157dB & 160dB & 164dB & 156dB & 164dB\\
Spectrum &
\begin{tabular}[t]{@{}l@{}}Unlicensed \textless1GHz\end{tabular} &
\begin{tabular}[t]{@{}l@{}}Unlicensed 900MHz\end{tabular} &
\begin{tabular}[t]{@{}l@{}l@{}}Licensed LTE\end{tabular} &
\begin{tabular}[t]{@{}l@{}}Licensed LTE\end{tabular} &
\begin{tabular}[t]{@{}l@{}}Licensed GSM\end{tabular} \\
Bandwidth & \textless500kHz & 100Hz &
\begin{tabular}[t]{@{}l@{}l@{}}180kHz (200KHz carrier)\end{tabular} &
\begin{tabular}[t]{@{}l@{}l@{}}1.08MHz (1.4MHz carrier)\end{tabular} & 200kHz \\
Data Rate & \textless50kbps & \textless100bps & \begin{tabular}[t]{@{}l@{}}\textless170kbps(DL),\textless250kbps(UL)\end{tabular} & \textless1Mbps & \textless140kbps \\
\bottomrule
\end{tabular}
\end{table*}

\section{Blockchain State of Affairs}
\label{sec_Blockchain}
\subsection{Blockchain}
\textit{Blockchain} is a distributed database deployed in a peer-to-peer network. \textit{Nodes} in the system create and broadcast \textit{transactions} continuously. Predictably, a blockchain consists of \textit{blocks}, which are cryptographically linked and timestamped collections of transactions. Nodes constantly verify blocks in the system to stand against malicious attackers trying to alter or forge transactions. All transactions in the system are signed using public-key cryptography and their authenticity is verifiable~\cite{nakamoto2008bitcoin}.

An in-depth look at the block structure of blockchain, as shown in Figure~\ref{fig:blockchain}, reveals that every block contains a block header and a varying number of transactions stored in a tree structure. In addition, every block header contains a timestamp and two hash values: one for a previous block\textquotesingle s header and another for all the transactions that are carried within that block. Because of this, it is possible to verify the integrity of the whole block, including all the transactions via block header.

\begin{minipage}{0.60\textwidth}
  \begin{figure}[H]
    \centering
    \includegraphics[scale=0.46]{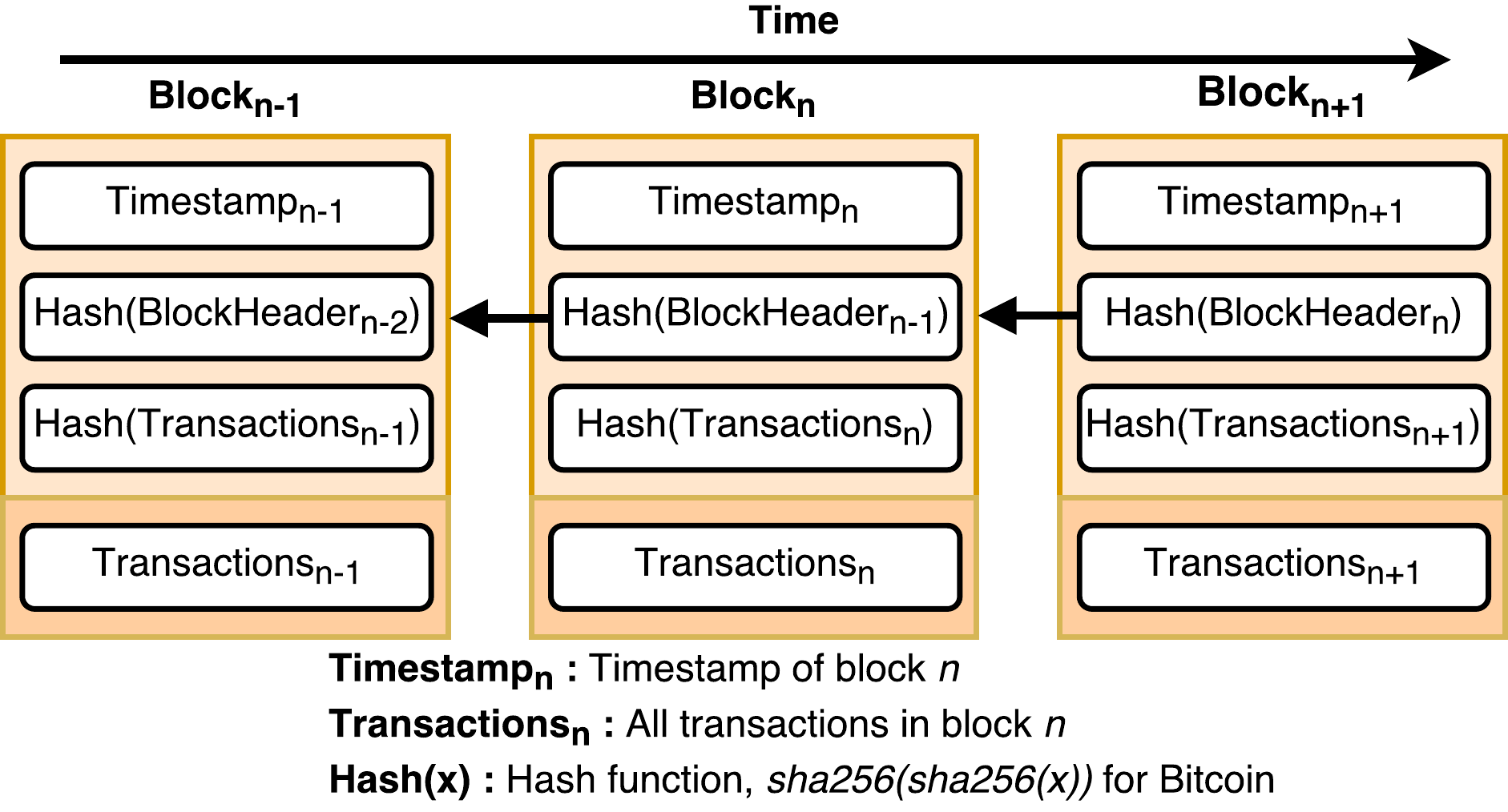}
    \caption{Blockchain Structure}
    \label{fig:blockchain}
  \end{figure}
\end{minipage}
\begin{minipage}{0.40\textwidth}
  \begin{figure}[H]
    \centering
    \includegraphics[scale=0.47]{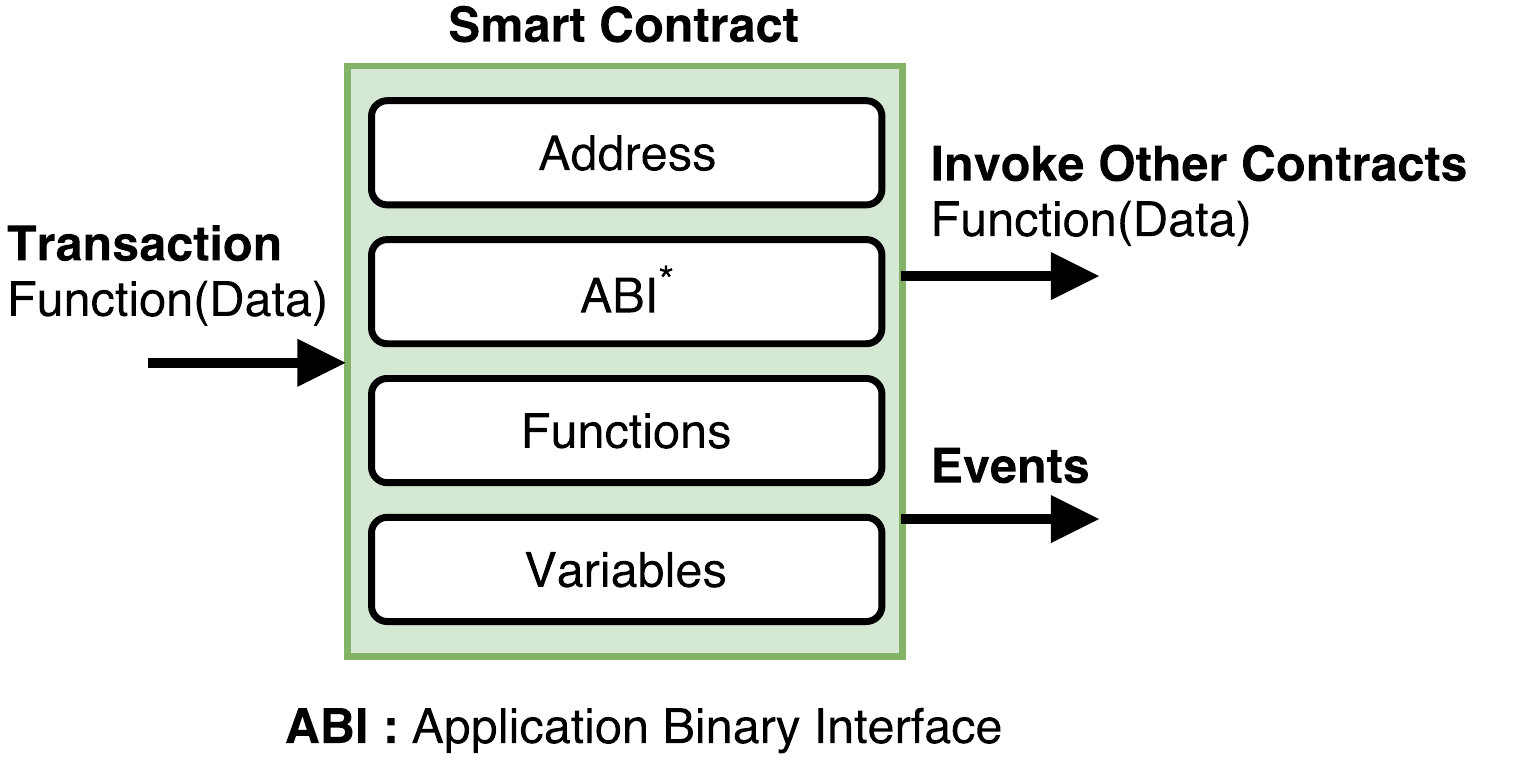}
    \caption{Smart Contract Structure}
    \label{fig:smart}
  \end{figure}
\end{minipage}

In traditional blockchain-based systems, special nodes called \textit{miners} try to find the next block by calculating a solution for a hard to compute, but easy to verify, mathematical problem, where the \textit{difficulty} of the problem is set as a constraint that is continuously changing. When a new block satisfying the current difficulty constraint is found, it\textquotesingle s propagated to the network as the next valid block and its miner is awarded for its efforts. A total block creation and propagation mechanism keeps all peers synchronized, i.e., in {\it consensus}. The difficulty of this mechanism is changed periodically to keep the block finding time in a predefined interval~\cite{nakamoto2008bitcoin}.

\subsection{Ethereum and Swarm}
\textit{Ethereum} is a blockchain-based infrastructure where stakeholders compile code fragments (\textit{smart contracts}) that may interact with each other or change the state of accounts on blockchain~\cite{ethereumwp}. Regular Bitcoin transactions contain sender and receiver addresses, value and a custom scripting system for verification. Ethereum extends the scripting capabilities of Bitcoin to a fully-fledged, Turing-complete programming language aiming to create a programming environment~\cite{ethereumwp}. As a result, Ethereum turns out to be a distributed application platform utilizing blockchain technology where users may pick arbitrary formats for transaction or ownership. Ethereum smart contracts are compiled virtual machine opcodes executed by the \textit{Ethereum Virtual Machine (EVM)}. The smart contract\textquotesingle s functions and events can only be accessed using the mined address of the contract and its application binary interface (ABI) (Figure~\ref{fig:smart}).

\textit{Swarm} is a peer-to-peer storage service that is DDOS-resistant, zero-downtime, fault-tolerant and censorship-resistant integrated with Ethereum~\cite{hartman1999swarm}. It is a torrent-like service with built-in incentives to guarantee uploaded data persistence due to high coupling with the Ethereum network layer. Hence, it is a strong candidate for a storage service targeting IoT.

\subsection{Consensus Algorithms}
\textbf{Proof-of-work (PoW)} is a consensus method, based on a hard to solve but easy to verify mathematical problem (one-way function) that is used by both Bitcoin and Ethereum. Calculation of acceptable hash values average 10 minutes and 17 seconds for Bitcoin and Ethereum respectively. With PoW algorithms, high computational power is needed to create even one block so that forging fake blocks or orchestrating a Sybil attack cannot occur.

\textbf{Proof-of-stake (PoS)} is another consensus approach, where the creator of the next block is chosen randomly. The randomness in selection is weighted by the amount of coins (i.e., stake) placed in the mechanism by the peer. PoS may increase protection against attacks as executing an attack will be expensive. Due to reduced computational work PoS requires less energy.

\textbf{Practical Byzantine Fault Tolerance (PBFT)} is a state-machine replication algorithm discovered to tolerate faults in distributed, low-latency storage systems \cite{castro1999practical}. Messages coming from the nodes are cryptographically signed and once enough identical responses are reached, consensus is met. However, unlike PoW or PoS, PBFT requires every node to know the entire set of its peer nodes participating in consensus~\cite{vukolic2015quest}. Although it is possible to have temporary blockchain forks because of network delays in PoW and PoS based systems, PBFT systems does not allow forks, satisfying consensus finality~\cite{vukolic2015quest}.

\subsection{Blockchain Node Types}
\textbf{Miner:} Miners are special nodes that pack transactions into blocks and run the consensus algorithms that satisfy system requirements to attain a financial benefit. In PoW consensus, miners in the network possess the highest computational power.

\textbf{Full node:} Full nodes download the whole blockchain and verify the integrity of all transactions continuously, making the infrastructure trustless and decentralized. Sufficient storage and computing power are required in order to run a full node.

\textbf{Thin client:} Thin clients only download the block headers that contain the hashes of the transactions within the block. Therefore, it is possible to interact with the blockchain with minimal storage and computing requirements. This approach is called \textit{Simplified Payment Verification (SPV)} in Bitcoin and \textit{Light Client} in Ethereum~\cite{nakamoto2008bitcoin}~\cite{ethereumwp}.

\textbf{Server-Trusting Client:} \textit{Bitcoin Client API (BCCAPI)} is proposed to make secure, light-weight clients for resource-constrained systems. With \textit{BCCAPI}, it is possible for a client to connect a server containing the blockchain and run queries against it. Here, the server has only public keys of clients and is unable to create a transaction without a client\textquotesingle s approval.

\section{IoT-Blockchain Integration Methods}
\label{sec_Integration}

Integrating IoT end devices and gateways to a blockchain infrastructure can be accomplished in many different ways depending on the capabilities and power requirements of end devices and gateway hardware. Assuming that end devices are either battery-powered or always-on and they are communicating with an always-on gateway connected to the internet (like a typical LPWAN case), one of the following integration strategies can be used for IoT gateways:

\textbf{Gateway as a full blockchain node:} IoT gateway operates as a full node, routing data to the network and verifying integrity at the same time. Integration is relatively easy because no changes are required in the way that end devices communicate. However, gateways should be powerful enough to operate as a full blockchain node. With the gateway\textquotesingle s total computing power for defending the integrity of the system, it is possible to achieve a trustless IoT infrastructure.

\textbf{Gateway as a thin client:} IoT gateway operates as a thin client by routing data to network and storing only relevant data fragments. Integration is relatively easy however the weakness here is that there should be other full nodes to defend the integrity of the system. A trustless infrastructure can still be achieved, but only with full nodes operating at the cloud side.

\begin{minipage}{0.44\textwidth}
\textbf{End devices as regular sensors:} Battery-powered end devices may be so weak that no additional client logic may be tolerated. In this case, no blockchain client is integrated. Transmitted data are received by an IoT gateway and are pushed to a blockchain infrastructure. This is suitable for extremely low-power sensors that do nothing more than broadcast their data.

\textbf{End devices as server-trusting client:} A blockchain client utilizing a BCCAPI-like interface may be integrated to battery-powered end devices. This way, the end device will interact with a blockchain node without any storage or computational requirements.

\textbf{End devices as thin client:} If end devices are not battery-powered and always on, they can operate as a thin client. Here, gateways can either be a full node or a transparent switch to relay transactions. If both gateways and end devices use blockchain clients, standardization in terms of data collection can be achieved.
\end{minipage}
\begin{minipage}{0.56\textwidth}
  \begin{figure}[H]
    \centering
    \includegraphics[scale=0.53]{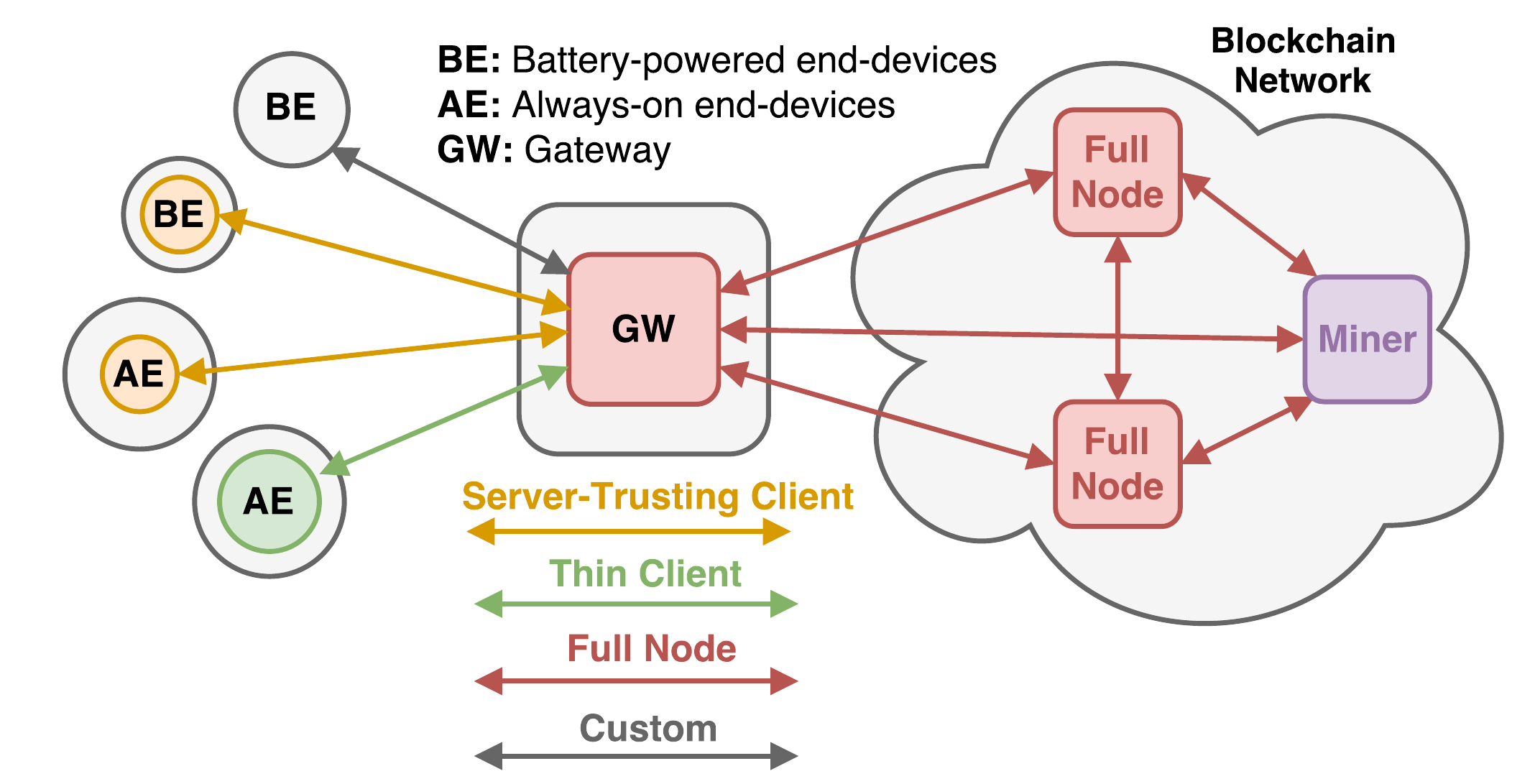}
    \caption{Blockchain Integration Methods~\cite{Ozyilmaz:2017:ILI:3125503.3125628}}
    \label{fig:integration}
  \end{figure}
  \vfill
\end{minipage}

\begin{table*}[htbp]
\centering
\caption{Roles and Communication Methods in LPWAN IoT Infrastructure}
\label{tab:integration}
\begin{tabular}{lllll}
\toprule
&
\begin{tabular}[t]{@{}l@{}}Battery-powered End Device\end{tabular} &
\begin{tabular}[t]{@{}l@{}}Always-on End Device\end{tabular} &
\begin{tabular}[t]{@{}l@{}}IoT Gateway\end{tabular} &
\begin{tabular}[t]{@{}l@{}}Cloud Backend\end{tabular}\\
\midrule
Traditional IoT &
\begin{tabular}[t]{@{}l@{}}Sensor with Custom Protocol\end{tabular} &
\begin{tabular}[t]{@{}l@{}}Sensor with Custom Protocol\end{tabular} &
\begin{tabular}[t]{@{}l@{}}Transparent Proxy\end{tabular} &
\begin{tabular}[t]{@{}l@{}}Centralized Core Services\end{tabular} \\  \rule{0pt}{3ex}
Blockchain IoT &
\begin{tabular}[t]{@{}l@{}l@{}l@{}}Server-trusting Client or\\Sensor with Custom Protocol\end{tabular} &
\begin{tabular}[t]{@{}l@{}l@{}}Thin Client or\\Server-trusting Client\end{tabular} &
\begin{tabular}[t]{@{}l@{}}Full Node or\\Thin Client\end{tabular} &
\begin{tabular}[t]{@{}l@{}}Miners and Full Nodes\end{tabular}\\
\bottomrule
\end{tabular}
\end{table*}

Table~\ref{tab:integration} shows the differences between traditional and blockchain-based IoT integration where every component acts as a part of a trustless peer-to-peer network and contributes to this network as much as its capabilities. This way, data collection and storage may be standardized by using blockchain client protocols. In a sample integration scenario (Figure~\ref{fig:integration}), gateways operate as full nodes and various end devices connect to it using different blockchain protocols.

\section{Proof of Concept}
\label{sec_PoC}
For the proof-of-concept implementation, LoRaWAN is selected due to being an unlicensed-band LPWAN technology with an affordable concentrator and end device hardware. In a previous implementation, a battery-powered LoRa end device\textquotesingle s position data was sent to a LoRa gateway, which then routed this data stream through the official Go-lang-based Ethereum client \textit{Geth} to a private Ethereum network using a smart contract~\cite{Ozyilmaz:2017:ILI:3125503.3125628}. This paper extends the preliminary work by a) storing IoT data not in blockchain but in the Swarm storage service, therefore eliminating the need for a private Ethereum network b) defining a clear way to access and retrieve data using Swarm and Ethereum smart contracts for additional applications like user interface services (e.g. data dashboard) or machine learning systems~\cite{bether}.

\begin{minipage}{0.40\textwidth}
A prototype system uses a LoRa end device, built with an RPi 2 connected to a Dragino LoRa/GPS Hat and a LoRa gateway, built with an RPi 3 connected to a LoRa concentrator board named iC880A from IMST. IoT gateway runs LoRa protocol software to communicate with low-power end devices. LoRa protocol software consists of a concentrator card driver and a network daemon to forward data packets into a local proxy server. This local server, called \textit{smart proxy}, receives data from the packet forwarder and acts as a mediator to push data into the blockchain-based infrastructure. Finally, Swarm and Ethereum clients complete the data flow. (Figure~\ref{fig:software})
\end{minipage}
\begin{minipage}{0.6\textwidth}
  \begin{figure}[H]
    \centering
    \includegraphics[scale=0.50]{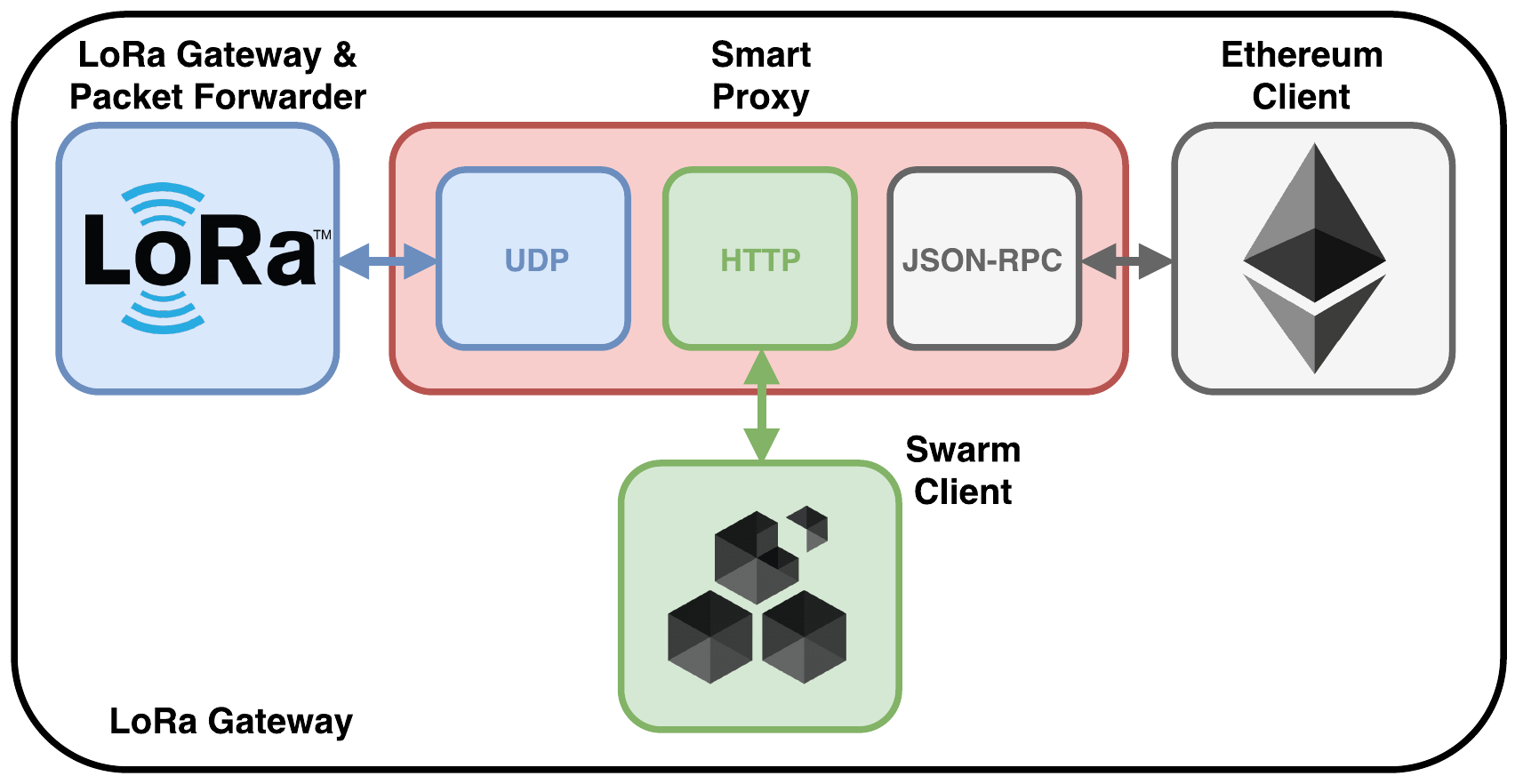}
    \caption{LoRa Gateway Software Architecture}
    \label{fig:software}
  \end{figure}
  \vfill
\end{minipage}

LoRa end devices wait for their turn and send their data without the need of establishing a connection to any specific party. Always-listening network daemon on the LoRa gateway picks up the transmission and forwards it to the smart proxy. After data is received, it is pushed into the Swarm file storage network using HTTP. A file hash is received and that same file hash will be used to access that particular data in the future (Figure~\ref{fig:uml}).

A smart proxy may communicate with the Ethereum client by means of its JSON-RPC interface, however, to enable a real interaction with the Ethereum network, a smart contract should be deployed first. After being compiled into bytecode, smart contracts are sent just like any other transaction, to be mined by miners. When a smart contract is mined, its address and application binary interface (ABI) are used to interact with it.

\begin{minipage}{0.5\textwidth}
\begin{listing}[H]
\begin{minted}[mathescape, linenos, numbersep=5pt, frame=lines, framesep=2mm, fontsize=\scriptsize]{csharp}
// IoT device data
struct device_data {
    // link for detecting devices
    uint index;
    // blockchain timestamps
    uint[] timestamps;
    // map timestamp values to swarm file hashes
    mapping(uint => string) filehashes;
}
// device id array of all received id's
address[] private device_index;
// map device id's to their data
mapping(address => device_data) private device_logs;
// event to log action
event log_action (address indexed device_id,
		  uint index,
		  uint timestamp,
		  string filehash);
\end{minted}
\caption{Smart Contract Data Structure}
\label{list:code}
\end{listing}
\end{minipage}
\begin{minipage}{0.5\textwidth}
\begin{listing}[H]
\begin{minted}[mathescape, linenos, numbersep=5pt, frame=lines, framesep=2mm, fontsize=\scriptsize]{csharp}
// set Swarm data handle for source device
function set_device_data (address device_id,
			  string filehash)
  public returns (uint index,
		  uint timestamp) {
  // get current block timestamp
  ts = now;
  // store data receive time (block timestamp)
  device_logs[device_id].timestamps.push(ts);
  // store swarm handle
  device_logs[device_id].filehashes[ts] = filehash;
  // store device link
  device_logs[device_id].index = device_index.push(device_id)-1;
  // trigger event, signalling received data
  log_action(device_id, device_index.length-1, ts, filehash);
  // return device index and timestamp
  return(device_index.length-1, ts);
}
\end{minted}
\caption{Smart Contract: Store Swarm File Hash}
\label{list:adddata}
\end{listing}
\end{minipage}

\begin{minipage}{0.45\textwidth}
Our smart contract code contains one event, which is \textit{log\_action}, and six functions named as \textit{is\_device\_present()}, \textit{get\_device\_count()}, \textit{get\_device\_at\_index()}, \textit{get\_device\_timestamps()}, \textit{get\_device\_data()} and \textit{set\_device\_data()}. All functions except \textit{set\_device\_data()} are \textit{constant} functions that do not change the contract state. Only \textit{set\_device\_data()} adds data to blockchain, thus a transaction should be set up. Listing~\ref{list:code} shows the actual code fragment declaring how device identifications, timestamp values and Swarm file hashes are connected. IoT gateways and their stored file hashes in the blockchain can be easily enumerated and accessed by using \textit{get\_device\_timestamp()} and \textit{get\_device\_data()} functions. \textit{set\_device\_data()} is the actual smart contract function (Listing~\ref{list:adddata}) that creates and maps a Swarm file hash to the current block timestamp. As soon as new data is added, \textit{log\_action} event is fired and all peers that are listening to that event get a callback. Ethereum smart contract and LoRa proxy code used for this implementation can be found on the 'Bether' project page~\cite{bether}.
\end{minipage}
\begin{minipage}{0.5\textwidth}
\begin{figure}[H]
  \centering
  \includegraphics[scale=0.78]{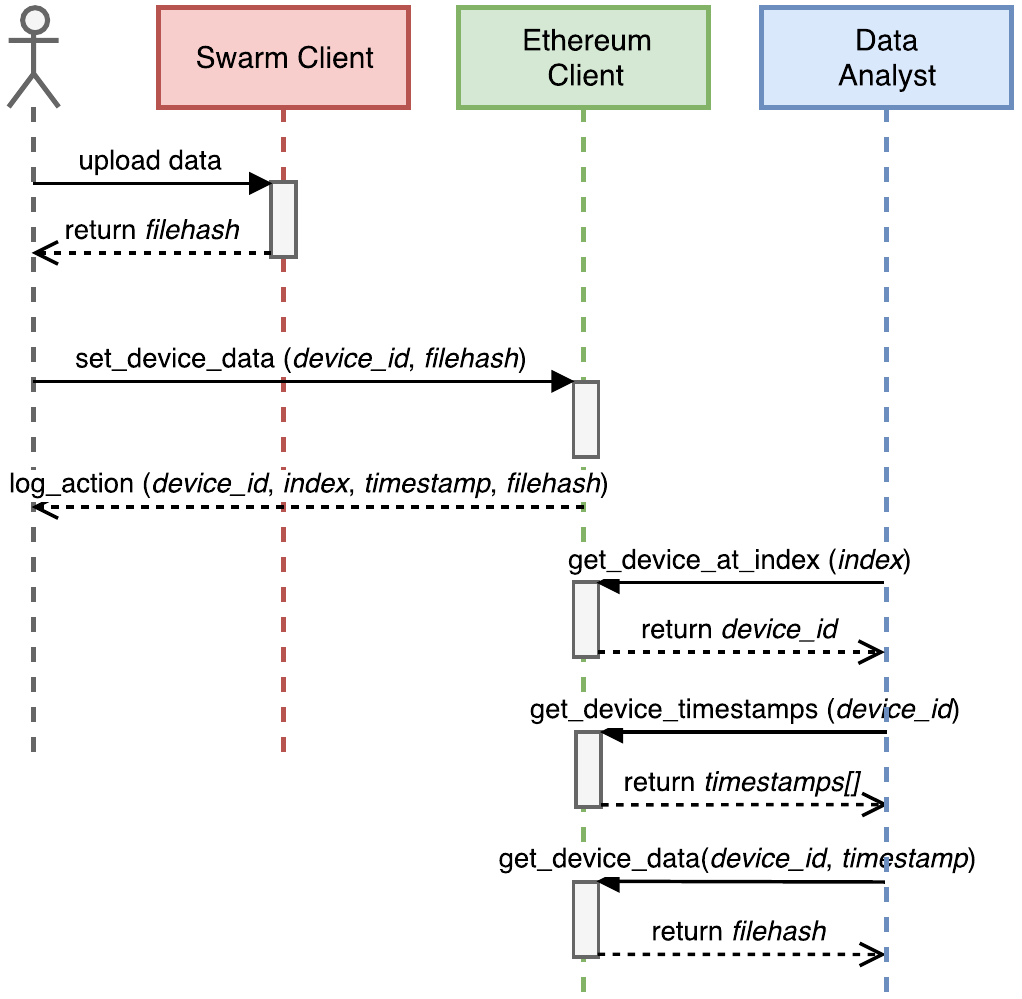}
  \caption{IoT Data Store and Access Scenario}
\label{fig:uml}
\end{figure}
  \vfill
\end{minipage}

\section{Evaluation}
\label{sec_Evaluation}

\subsection{Resource consumption}
There are at least three different types of Ethereum node configurations in our setup to accommodate underlying hardware resources. The usage statistics below are given, a) for a one-month-old, private Ethereum installation (around 200k blocks), b) for the public Ethereum blockchain. Peak memory consumption may vary from setup to setup due to synchronization speed:

\textbf{Mining Full Node:} With an active Swarm client, these nodes use between 1.2GB and 1.5GB of memory in private Ethereum network. In public network they need at least 4GB of memory; it is expected that this requirement will go up in time. These nodes are powerful servers deployed in the cloud.

\textbf{Non-mining Full Node:} With an active Swarm client, these nodes use between 300MB and 400MB of memory in private Ethereum network. In public network they need at least 2GB of memory to properly sync with blockchain. These nodes may be IoT vendor servers, network provider servers or powerful IoT gateways.

\textbf{Non-mining Light Node:} With an active Swarm client, these nodes use around 300MB of memory, whereas Ethereum client only uses around 50MB (200MB in public Ethereum). Because Swarm has no light client mode to limit bandwidth or memory usage at the moment, the memory benefit is minimal. These nodes may be regular IoT gateways and end devices.

\subsection{Data Throughput}
Data throughput in blockchain systems depends on various metrics and varies in different implementations. Bitcoin imposes throughput limits with its 10-minute average block time and fixed 1MB block size. In Ethereum\textquotesingle s case, there is no fixed block size but a gas limit per block (i.e., amount of resource to be used by transactions). Similar to resource statistics, throughput statistics are given, a) for one-month-old, private Ethereum installation, b) for the public Ethereum blockchain.

Our private Ethereum network has a gas limit of 4,712,388 gas/block and the average gas price is 21k gas, therefore, a block may only contain 224 transactions. Considering that the average block time is 14 seconds for the private system, the throughput will be 16 transactions per second (or around 1k transactions per minute). The public Ethereum is in the middle of a difficulty increase as of September 2017. At the time of writing, the public Ethereum system has a gas limit of 6,718,904 gas/block, an average gas price of 21k gas and an average block time of 30 seconds. Data throughput will in turn be 320 transactions per block, which is 10.6 transactions per second (or 640 transactions per minute).

Though transaction throughput seems low to support a full-scale deployment today, it is imperative to note that transactions are created only by IoT gateways and every gateway may serve hundreds of thousands of end devices. Proposed infrastructure can support tens of thousands of IoT gateways (and millions of end devices) pushing data periodically every 15 minutes.

\section{Discussion}
\label{sec_Discussion}
This section deals with which parts of blockchain systems may be improved for a better IoT integration.

\textbf{Inefficiency:} Bitcoin and Ethereum use PoW algorithms that guarantee every mined block is backed by a certain amount of computational work. This approach is inherently inefficient because every miner in the system is doing hard calculations individually. In the IoT blockchain, a PoS based consensus may be much more suitable as discussed in Section~\ref{sec_Blockchain}. PoS algorithms may create monopolies due to concentration of stake, but in the case of IoT, this "bug" may be used as a "feature". Empowering certain trusted parties like system integrators or regulators may indeed be beneficial.

\textbf{Encryption and Access Control:} Blockchain based systems store clear data although transactions are signed with public-key crypto. When an IoT system is dealing with sensitive data, either payload must be encrypted before pushed into blockchain, or sophisticated mechanisms to conceal critical data should be used. As an example Zcash payments are published on a public blockchain, but the sender, recipient, and the amount of a transaction remain private by utilizing zero-knowledge proofs~\cite{sasson2014zerocash}.

\textbf{Bandwidth:} Gateways in LPWANs are the point of transmission to the cloud for connected end devices. All LPWANs consider that gateways to be connected to a fast communication link, either wired or wireless. If gateways operate as a full node, bandwidth requirements will increase considerably because of the messaging and synchronization traffic of blockchain.

\textbf{Real-Time Systems:} Due to their trustless nature, blockchain-based systems may be able to store data only after a certain period of time, which is determined by the block creation interval. In order to support real-time applications, data propagation delay should be minimized by proposing new types of consensus functions that are fine-tuned for IoT scenarios.

\section{Conclusion}
Internet of Things is the key to smarter cities, transportation systems, energy systems, and healthcare. In order to deal with the increasing number of IoT devices, it is necessary to standardize the method of communication for IoT gateways and create a common IoT backend. Using blockchain\textquotesingle s decentralized, trustless nature in combination with DDOS-resistant, fault-tolerant data storage, a new type of IoT backend may be created. In this way, all kinds of IoT end devices may be integrated to this infrastructure based on their computing and storage capabilities. Such an achievement will lead to data-centric business models where application development and data processing can be massively conducted by using smart contracts as demonstrated with our proof-of-concept Bether~\cite{bether}.

\section{About the Authors}
\textbf{Kazım Rıfat Özyılmaz} (kazim@monolytic.com) earned his B.Sc. degree in Electronics and Communication Engineering from Istanbul Technical University in 2004 and his M.Sc. degree in Software Engineering from Bogazici University in 2015. He is a Ph.D. student in Computer Engineering at Bogazici University, working on blockchain technology and IoT integration.

\textbf{Arda Yurdakul} (yurdakul@boun.edu.tr) is a Professor in Computer Engineering, Bogazici University, Istanbul, Turkey. Her research interests include embedded systems, IoT, blockchain technology, real-time systems, reconfigurable computing.

\bibliographystyle{IEEEtran}
\bibliography{IEEEabrv,cesoc}

\begin{thebibliography}{10}
\providecommand{\url}[1]{#1}
\csname url@samestyle\endcsname
\providecommand{\newblock}{\relax}
\providecommand{\bibinfo}[2]{#2}
\providecommand{\BIBentrySTDinterwordspacing}{\spaceskip=0pt\relax}
\providecommand{\BIBentryALTinterwordstretchfactor}{4}
\providecommand{\BIBentryALTinterwordspacing}{\spaceskip=\fontdimen2\font plus
\BIBentryALTinterwordstretchfactor\fontdimen3\font minus
  \fontdimen4\font\relax}
\providecommand{\BIBforeignlanguage}[2]{{%
\expandafter\ifx\csname l@#1\endcsname\relax
\typeout{** WARNING: IEEEtran.bst: No hyphenation pattern has been}%
\typeout{** loaded for the language `#1'. Using the pattern for}%
\typeout{** the default language instead.}%
\else
\language=\csname l@#1\endcsname
\fi
#2}}
\providecommand{\BIBdecl}{\relax}
\BIBdecl

\bibitem{mohanty2016everything}
S.~P. Mohanty, U.~Choppali, and E.~Kougianos, ``Everything you wanted to know
  about smart cities: The internet of things is the backbone,'' \emph{IEEE
  Consumer Electronics Magazine}, vol.~5, no.~3, pp. 60--70, 2016.

\bibitem{montgomery2015future}
B.~Montgomery, ``{Future Shock: IoT benefits beyond traffic and lighting energy
  optimization},'' \emph{IEEE Consumer Electronics Magazine}, vol.~4, no.~4,
  pp. 98--100, 2015.

\bibitem{lee2017bidaas}
J.~Lee, ``{BIDaaS: Blockchain based ID as a Service},'' \emph{IEEE Access},
  2017.

\bibitem{zorunluluk}
{D. Puthal, N. Malik, S. P. Mohanty, E. Kougianos, and C. Yang}, ``{The
  Blockchain as a Decentralized Security Framework},'' \emph{IEEE Consumer
  Electronics Magazine}, vol.~7, no.~2, pp. 18--21, 2018.

\bibitem{corcoran2016mobile}
P.~Corcoran and S.~K. Datta, ``{Mobile-edge computing and the internet of
  things for consumers: Extending cloud computing and services to the edge of
  the network},'' \emph{IEEE Consumer Electronics Magazine}, vol.~5, no.~4, pp.
  73--74, 2016.

\bibitem{gubbi2013internet}
J.~Gubbi, R.~Buyya, S.~Marusic, and M.~Palaniswami, ``{Internet of Things
  (IoT): A vision, architectural elements, and future directions},''
  \emph{Future generation computer systems}, vol.~29, no.~7, pp. 1645--1660,
  2013.

\bibitem{ericsson2016cellular}
\BIBentryALTinterwordspacing
Ericsson. (2016) Cellular networks for massive {IoT}. [Online]. Available:
  \url{https://www.ericsson.com/res/docs/whitepapers/wp_iot.pdf}
\BIBentrySTDinterwordspacing

\bibitem{vangelista2015long}
L.~Vangelista, A.~Zanella, and M.~Zorzi, ``{Long-range IoT technologies: The
  dawn of LoRa},'' in \emph{Future Access Enablers of Ubiquitous and
  Intelligent Infrastructures}.\hskip 1em plus 0.5em minus 0.4em\relax
  Springer, 2015, pp. 51--58.

\bibitem{nakamoto2008bitcoin}
\BIBentryALTinterwordspacing
S.~Nakamoto. (2008) Bitcoin: A peer-to-peer electronic cash system. [Online].
  Available: \url{https://bitcoin.org/bitcoin.pdf}
\BIBentrySTDinterwordspacing

\bibitem{DBLP:journals/corr/EyalS13}
\BIBentryALTinterwordspacing
I.~Eyal and E.~G. Sirer, ``Majority is not enough: Bitcoin mining is
  vulnerable,'' \emph{CoRR}, vol. abs/1311.0243, 2013. [Online]. Available:
  \url{http://arxiv.org/abs/1311.0243}
\BIBentrySTDinterwordspacing

\bibitem{3gpp2016progress}
\BIBentryALTinterwordspacing
3GPP. (2016) Progress on {3GPP IoT}. [Online]. Available:
  \url{http://www.3gpp.org/news-events/3gpp-news/1766-iot_progress}
\BIBentrySTDinterwordspacing

\bibitem{bardyn2016iot}
J.~Bardyn, T.~Melly, O.~Seller, and N.~Sornin, ``{IoT: The era of LPWAN is
  starting now},'' in \emph{European Solid-State Circuits Conference, ESSCIRC
  Conference 2016: 42nd}.\hskip 1em plus 0.5em minus 0.4em\relax IEEE, 2016,
  pp. 25--30.

\bibitem{nokia2016lteevo}
\BIBentryALTinterwordspacing
Nokia. (2015) {LTE} evolution for {IoT} connectivity. [Online]. Available:
  \url{http://resources.alcatel-lucent.com/asset/200178}
\BIBentrySTDinterwordspacing

\bibitem{ethereumwp}
\BIBentryALTinterwordspacing
V.~Buterin. (2014) A next-generation smart contract and decentralized
  application platform. [Online]. Available:
  \url{https://github.com/ethereum/wiki/wiki/White-Paper}
\BIBentrySTDinterwordspacing

\bibitem{hartman1999swarm}
J.~H. Hartman, I.~Murdock, and T.~Spalink, ``The swarm scalable storage
  system,'' in \emph{Distributed Computing Systems, 1999. Proceedings. 19th
  IEEE International Conference on}.\hskip 1em plus 0.5em minus 0.4em\relax
  IEEE, 1999, pp. 74--81.

\bibitem{castro1999practical}
M.~Castro and B.~Liskov, ``Practical byzantine fault tolerance,'' in
  \emph{OSDI}, vol.~99, 1999, pp. 173--186.

\bibitem{vukolic2015quest}
M.~Vukoli{\'c}, ``The quest for scalable blockchain fabric: Proof-of-work vs.
  bft replication,'' in \emph{International Workshop on Open Problems in
  Network Security}.\hskip 1em plus 0.5em minus 0.4em\relax Springer, 2015, pp.
  112--125.

\bibitem{Ozyilmaz:2017:ILI:3125503.3125628}
\BIBentryALTinterwordspacing
K.~R. \"{O}zy{\i}lmaz and A.~Yurdakul, ``{Integrating Low-power IoT Devices to
  a Blockchain-based Infrastructure: Work-in-progress},'' in \emph{Proceedings
  of the Thirteenth ACM International Conference on Embedded Software 2017
  Companion}, ser. EMSOFT '17.\hskip 1em plus 0.5em minus 0.4em\relax ACM,
  2017, pp. 13:1--13:2. [Online]. Available:
  \url{http://doi.acm.org/10.1145/3125503.3125628}
\BIBentrySTDinterwordspacing

\bibitem{bether}
\BIBentryALTinterwordspacing
K.~R. {\"{O}zy{\i}lmaz}. (2017) Bether: {IoT} backend with {Swarm} and
  {Ethereum} smart contracts. [Online]. Available:
  \url{https://github.com/kozyilmaz/bether}
\BIBentrySTDinterwordspacing

\bibitem{sasson2014zerocash}
E.~B. Sasson, A.~Chiesa, C.~Garman, M.~Green, I.~Miers, E.~Tromer, and
  M.~Virza, ``Zerocash: Decentralized anonymous payments from bitcoin,'' in
  \emph{Security and Privacy (SP), 2014 IEEE Symposium on}.\hskip 1em plus
  0.5em minus 0.4em\relax IEEE, 2014, pp. 459--474.

\end{thebibliography}

\end{document}